\documentclass[12pt]{article}
\usepackage{amssymb,amsmath,amsfonts}
\usepackage{graphicx}
\usepackage{graphics}
\usepackage{eepic,epsfig}
\usepackage{dcolumn}


\begin{document}

\title{Intense radiation from a relativistic electron rotating about a
dielectric ball}
\author{M.L. Grigoryan\thanks{%
E-mail address: levonshg@web.am} \\
\textit{Institute of Applied Problems in Physics}\\
\textit{25 Hr. Nersessian Str., 375014 Yerevan, Armenia}\\
}
\date{\today }
\maketitle

\begin{abstract}
The radiation from a relativistic electron uniformly rotating along an orbit
in the equatorial plane of a dielectric ball was calculated taking into
account the dielectric losses of energy and dispersion of electromagnetic
oscillations inside the substance of ball. It was shown that due to the
presence of ball the radiation from the particle at some harmonics may be
several dozens of times more intense than that from the particle rotating in
an infinite homogeneous (and transparent) dielectric. The generation of such
a high power radiation is possible only at some particular values of the
ratio of ball radius to that of electron orbit and when the Cherenkov
condition for the ball material and the velocity of particle "image" on the
ball surface is met.
\end{abstract}



\bigskip

\section{Introduction}

\label{section1}

Due to such unique properties as the high intensity, high degree of
collimation, and wide spectral range (see \cite{1}-\cite{6} and references
therein) Synchrotron Radiation (SR) serves as an extraordinary research tool
for advanced studies in both the fundamental and applied sciences. These
applications motivate the importance of analyzing various controlling
mechanisms of SR parameters. From this point of view it is of interest to
study the influence of medium on SR.

The characteristics of high-energy electromagnetic processes in the presence
of material are essentially changed by giving rise to new phenomena such as
Cherenkov Radiation (CR) \cite{7}-\cite{9}, transition radiation \cite{10,11}%
, parametric X-ray radiation, channeling radiation etc. The operation of a
number of devices intended for production of electromagnetic radiation is
based on interactions of relativistic electrons with matter (see, e.g., \cite%
{12}).

In \cite{13} (see also \cite{14,9}) it was shown based on the consideration
of SR from a charged particle circulating in a homogeneous medium, that the
interference between SR and CR had interesting consequences. New interesting
phenomena occur in case of inhomogeneous media. A well-known example here is
the transition radiation. In particular, the interfaces between media can be
used for monitoring the flows of radiation emitted in various systems. In a
series of papers initiated in \cite{15}-\cite{17} the cases of simplest
geometry boundaries, namely, the boundaries with spherical \cite{15,16,18,19}
and cylindrical \cite{17},\cite{20}-\cite{24} symmetries, have been
investigated. As a result, nontrivial peculiarities of CR induced at such
boundaries were revealed. E.g., in \cite{18,19} the spectral distribution of
radiation intensity from a relativistic particle uniformly rotating about
(or inside) a dielectric ball, in its equatorial plane was investigated. It
was shown that under certain conditions the presence of ball leads to an
interesting effect, - at a definite frequency (harmonic) the rotating
particle radiates energy greater by few orders of magnitude than that in a
continuous and unbounded dielectric. However, in \cite{18,19} the phenomena
of absorption and dispersion of electromagnetic waves were not taken into
account. The present paper is meant to fill that gap.

It is organized as follows: in Section \ref{section2} the description of
problem is given, in Section \ref{section3} the method of solution is
described and the final analytical expression for intensity of radiation
from a relativistic particle rotating about a ball is given. Numerical
results are presented in Section \ref{section4} and the main results are
summarized in the last section of paper.

\section{The description of the problem}

\label{section2}

Now consider a relativistic particle (electron), which uniformly rotates in
a magnetic field around a dielectric ball, in its equatorial plane. We shall
restrict ourselves to the simplest case, when the space outside of the ball
is empty. In the case under consideration the permittivity $\qquad \qquad
\qquad \qquad \qquad \qquad \qquad \qquad $%
\begin{equation}
\varepsilon \left( r\right) =\varepsilon _{b}+\left( 1-\varepsilon
_{b}\right) \Theta \left( r-r_{b}\right) ,  \label{2.1}
\end{equation}%
where $r_{b}$ is the ball radius, $\varepsilon _{b}=\varepsilon
_{b}^{^{\prime }}+i\varepsilon _{b}^{^{\prime \prime }}$ \ is the
permittivity of ball material (in general, it is a complex-valued quantity),
and $\Theta $\ is a unit step function (the origin of a spherical
coordinates system is located at the center of ball). The density of
electrical current
\begin{equation}
\mathbf{j}(\mathbf{r},t)=\frac{e\text{v}}{r_{e}^{2}}\mathbf{e}_{\varphi
}\delta (r-r_{e})\delta (\theta -\frac{\pi }{2})\delta (\varphi -\omega
_{0}t),  \label{2.2}
\end{equation}%
where $r_{e}$ is the electron orbit radius ($r_{e}>r_{b}$), and $\mathrm{v}%
=r_{e}\omega _{0}$ is the linear velocity of electron, $\theta =\pi /2$\
corresponds to the equatorial plane of ball.

The uniform rotation of electron is accompanied by radiation at discrete
frequencies (harmonics)%
\begin{equation}
\omega _{k}=k\omega _{0},  \label{2.3}
\end{equation}%
where $k=1;2;3...$\ We assume that the electron braking due to the radiation
is compensated by an external (e.g., electrical) force driving the particle
to move uniformly around a circle. The intensity $I_{k}$ of radiation
averaged over a gyration period is determined by the expression \cite{25}%
\begin{equation}
I_{k}=\frac{c}{2\pi }\underset{r\rightarrow \infty }{\lim }r^{2}\int
\left\vert \mathbf{\nabla }\times \mathbf{A}(\mathbf{r},\omega _{k}\mathbf{)}%
\right\vert ^{2}d\Omega  \label{2.4}
\end{equation}%
($\Omega $ is the spatial angle). The Fourier transform%
\begin{equation}
\mathbf{A(r},\omega _{k})=\frac{1}{T}\int_{0}^{T}\mathbf{A[r,}t]\exp
(ik\omega _{0}t)dt  \label{2.5}
\end{equation}%
of the vector potential is determined by equation%
\begin{equation}
(\Delta +\frac{\omega _{k}^{2}}{c^{2}}\varepsilon )\mathbf{A-}\frac{1}{%
\varepsilon }(\mathbf{\nabla }\varepsilon )\mathbf{\nabla }\cdot \mathbf{A=-}%
\frac{4\pi }{c}\mathbf{j}  \label{2.6}
\end{equation}%
(see, e.g., \cite{10}, the permeability of substance being assumed to be 1)
under Lorenz gauge condition:%
\begin{equation}
\mathbf{\nabla }\cdot \mathbf{A}-i\varepsilon \frac{\omega _{k}}{c}\psi =0,
\label{2.7}
\end{equation}%
where $\psi (\mathbf{r},\omega _{k})$ is the time component of
electromagnetic field 4-potential.

It is convenient to introduce a dimensionless quantity%
\begin{equation}
TI_{k}/\hbar \omega _{k}\equiv n_{k},  \label{2.8}
\end{equation}%
where $TI_{k}$\ is the energy radiated at frequency $\omega _{k}$ during one
period of electron rotation, and $\hbar \omega _{k}\ $is the energy of
corresponding quantum of electromagnetic wave. As a result, the total energy
$W_{T}$ radiated in time $T$ is determined by the expression%
\begin{equation}
W_{T}=\sum_{k=1}^{\infty }n_{k}\hbar \omega _{k}.  \label{2.9}
\end{equation}

If all space is filled with transparent material with real and constant $%
\varepsilon $, then \cite{9}%
\begin{equation}
n_{k}\left( \infty ;\mathrm{v},\varepsilon \right) =\dfrac{n_{0}}{\beta
\sqrt{\varepsilon }}\left[ 2\beta ^{2}J_{2k}^{^{\prime }}\left( 2k\beta
\right) +\left( \beta ^{2}-1\right) \int_{0}^{2k\beta }J_{2k}\left( x\right)
dx\right] ,  \label{2.10}
\end{equation}%
where%
\begin{equation}
n_{0}=2\pi e^{2}/\hbar c\cong 0.0459,\qquad \beta =\text{v}\sqrt{\varepsilon
}/c,  \label{2.11}
\end{equation}%
and $J_{k}\left( x\right) $\ is the Bessel function of integer order. In
this formula the\ case of $\varepsilon =1$ corresponds to synchrotron
radiation in vacuum (see, e.g., \cite{2,25}).

We aim at the solution of equation (\ref{2.6}) subject to condition that $%
\varepsilon (r)$\ is determined by equation (\ref{2.1}). Then, using (\ref%
{2.4}) and (\ref{2.8}), we shall try to find the values of $r_{b}$\ and $%
\varepsilon _{b}$ (parameters of the ball), at which the number $n_{k}\left(
\text{ball};\text{v},r_{b}/r_{e},\varepsilon _{b}\right) $\ of quanta
generated by the rotating particle is maximum.

\section{The method of calculations and the final formula}

\label{section3}

An arbitrary vector field $\mathbf{S}$ can be expanded in terms of spherical
vectors:%
\begin{equation}
\mathbf{S}(\mathbf{r})=\sum_{\mu =1}^{3}\sum_{l=0}^{\infty
}\sum_{m=-l}^{l}S_{\mu }^{lm}(r)\mathbf{X}_{lm}^{(\mu )}(\Omega ),
\label{3.1}
\end{equation}%
where%
\begin{equation}
S_{\mu }^{lm}(r)=\int \mathbf{X}_{lm}^{(\mu )\ast }\cdot \mathbf{S}(\mathbf{r%
})d\Omega ,  \label{3.2}
\end{equation}%
and $\mathbf{X}_{lm}^{(\mu )}$ are spherical vectors of longitudinal ($\mu
=1 $), electric ($\mu =2$ ) and magnetic ($\mu =3$) types \cite{26}:%
\begin{eqnarray}
\mathbf{X}_{lm}^{(1)} &=&\mathbf{n}Y_{lm}(\Omega ),\quad -l\leq m\leq
l,\quad l=0,\ 1,\ 2...,\quad \mathbf{n}=\mathbf{r}/r  \notag \\
\mathbf{X}_{lm}^{(2)} &=&\frac{\mathbf{\nabla }_{\mathbf{n}}Y_{lm}}{\sqrt{%
l(l+1)}},\quad \mathbf{X}_{lm}^{(3)}=\frac{\mathbf{n\times \nabla }_{\mathbf{%
n}}}{\sqrt{l(l+1)}}Y_{lm},\quad \mathbf{X}_{00}^{(2)}=\mathbf{X}%
_{00}^{(3)}=0.  \label{3.3}
\end{eqnarray}%
The operator%
\begin{equation}
\mathbf{\nabla }_{\mathbf{n}}=r(\mathbf{\nabla }-\mathbf{n}\partial
/\partial r)=\mathbf{e}_{\theta }\frac{\partial }{\partial \theta }+\frac{%
\mathbf{e}_{\varphi }}{\sin \theta }\frac{\partial }{\partial \varphi }
\label{3.4}
\end{equation}%
acts on the $\mathbf{n}$-dependent functions, and $Y_{lm}(\Omega )$\ are
spherical harmonics satisfying the equation%
\begin{equation}
\Delta _{\mathbf{n}}Y_{lm}=-l(l+1)Y_{lm},  \label{3.5}
\end{equation}%
where $\Delta _{\mathbf{n}}=\mathbf{\nabla }_{n}\cdot \mathbf{\nabla }_{n}$\
is the angular part of Laplacian.

According to (\ref{3.1}), the determination of $\mathbf{S}(\mathbf{r})=%
\mathbf{A}(\mathbf{r})$\ reduces to finding the quantity $A_{\mu }^{lm}(r)$
that is independent of $\Omega =(\theta ,\varphi )$\ (for simplicity here
and in what follows $\omega _{k}$ is omitted in $\mathbf{S}(\mathbf{r,}%
\omega _{k})$\ and analogous quantities). The calculation of $A_{\mu
}^{lm}(r)$ is simplified due to the spherical symmetry of ball. Really,
\begin{equation}
A_{\mu }^{lm}(r)=\sum_{\nu =1}^{3}\sum_{l_{1}=0}^{\infty
}\sum_{m_{1}=-l_{1}}^{l_{1}}\int_{0}^{\infty }G_{\mu \nu }\left(
r,r^{^{\prime }};lm,l_{1}m_{1}\right) j_{\nu }^{l_{1}m_{1}}(r^{^{\prime
}})dr^{^{\prime }}.  \label{3.6}
\end{equation}%
Such a writing of the solution of equation (\ref{2.6}) is convenient because
it permits to find a radiation field for an arbitrary current $\mathbf{j}$,
provided that the Green Function\ (GF) $G_{\mu \nu }\ $of electromagnetic
field is known. By direct substitution of (\ref{3.6}), (\ref{3.1}) and (\ref%
{2.1}) into (\ref{2.6}) one can make sure that for spherically symmetric
medium
\begin{equation}
G_{\mu \nu }\left( r,r^{^{\prime }};lm,l_{1}m_{1}\right) =\delta
_{ll_{1}}\delta _{mm_{1}}G_{\mu \nu }\left( r,r^{^{\prime }};l\right)
\label{3.7}
\end{equation}%
(GF is "dioganal" in $l$, $m$ and does not depend on $m$). Thus, in our case
$G_{\mu \nu }$\ is a 3x3 matrix depending only on $r$, $r^{^{\prime }}$ and $%
l$. This fact considerably simplifies the calculations. The problem is
aggravated by the fact that in (\ref{2.6}) the delta function%
\begin{equation}
\frac{\partial \varepsilon }{\partial r}=(1-\varepsilon _{b})\delta (1-r_{b})
\label{3.8}
\end{equation}%
is entered. A rather simple and visual method of GF evaluation in similar
situations may be obtained in \cite{27} (see also \cite{28}). The efficiency
of method is due to the transition from the differential equation for GF to
a relevant Lipmann-Schwinger type integral equation. Owing to the presence
of delta function in (\ref{3.8}), the Lipmann-Schwinger type integral
equation is transformed into an easily solved algebraic equation, and the
problem of evaluation is reduced to the solution of an "auxiliary"
differential equation without the delta function. In \cite{15}, GF of an
electromagnetic field in the medium consisting of an arbitrary number of
spherically symmetric layers with common center and various permittivities
was determined by means of this same method. Also in \cite{15} a
corresponding formula for intensity of radiation from a charged particle
arbitrarily moving in a layered spherically symmetric medium was derived.
The developed method was verified for known particular cases of the motion
of charged particle (i) in a homogeneous medium and (ii) through a flat
boundary between two homogeneous media (transition radiation). In \cite%
{16,18} the problem on radiation from a particle uniformly rotating about a
dielectric ball in its equatorial plane is solved. In particular cases the
derived analytical expressions coincide with the results known earlier \cite%
{9,25}. The radiation from an electron uniformly rotating along an orbit in
the equatorial plane inside dielectric ball was studied in \cite{19}.

In the absence of absorption and ionization losses, the work of external
force resisting the braking of particle motion should be equal to the
radiated energy. The radiated energy was calculated also by means of this
method \cite{18,19}. The coincidence of obtained results served as an
indirect confirmation of the correctness of calculations.

Below we give only the final formula for the number of quanta generated from
a relativistic particle during one revolution about a dielectric ball:%
\begin{equation}
n_{k}\left( \text{ball};\text{v},x,\varepsilon _{b}\right) =\dfrac{2n_{0}}{k}%
\sum_{s=0}^{\infty }\left( \left\vert a_{kE}\left( s\right) \right\vert
^{2}+\left\vert a_{kH}\left( s\right) \right\vert ^{2}\right)  \label{3.9}
\end{equation}%
Here $x\equiv r_{b}/r_{e}$, and%
\begin{eqnarray}
a_{kE}\left( s\right) &=&kb_{l}\left( E\right) P_{l}^{k}\left( 0\right)
\sqrt{\left( l-k\right) !/l\left( l+1\right) \left( 2l+1\right) \left(
l+k\right) !},\quad l=k+2s  \notag \\
a_{kH}\left( s\right) &=&b_{l}\left. \left( H\right) \sqrt{\dfrac{\left(
2l+1\right) \left( l-k\right) !}{l\left( l+1\right) \left( l+k\right) !}}%
\dfrac{dP_{l}^{k}\left( y\right) }{dy}\right\vert _{y=0},\quad l=k+2s+1
\label{3.10}
\end{eqnarray}%
are dimensionless amplitudes describing the contributions of multipoles of
electric ($E$) and magnetic ($H$) types respectively. In (\ref{3.10}) $%
P_{l}^{k}(y)$\ are the associated Legendre polynomials, and $b_{l}\left(
E\right) ,$ $b_{l}\left( H\right) $ are the following factors depending on $%
k $, v, $x$ and $\varepsilon _{b}$:
\begin{eqnarray}
b_{l}\left( H\right) &=&iu\left[ j_{l}\left( u\right) -h_{l}\left( u\right)
\dfrac{\left\{ j_{l}\left( xu_{b}\right) ,j_{l}\left( xu\right) \right\} _{x}%
}{\left\{ j_{l}\left( xu_{b}\right) ,h_{l}\left( xu\right) \right\} _{x}}%
\right] ,\quad u=k\text{v}/c,\quad u_{b}=k\text{v}\sqrt{\varepsilon _{b}}/c
\notag \\
b_{l}\left( E\right) &=&\left( l+1\right) b_{l-1}\left( H\right)
-lb_{l+1}\left( H\right) +x^{-2}\left( 1-\varepsilon _{b}\right) \left[ j_{%
\underline{l-1}}\left( xu_{b}\right) +j_{\underline{l+1}}\left(
xu_{b}\right) \right] \times  \notag \\
&&\times \left[ h_{\underline{l-1}}\left( u\right) +h_{\underline{l+1}%
}\left( u\right) \right] \frac{l\left( l+1\right) u_{b}j_{l}\left(
xu_{b}\right) }{lz_{l-1}^{l}+\left( l+1\right) z_{l+1}^{l}},  \label{3.11}
\end{eqnarray}%
where $h_{l}\left( y\right) =j_{l}\left( y\right) +i$n$_{l}\left( y\right) $%
, and $j_{l}\left( y\right) $, \ n$_{l}\left( y\right) $ are spherical
Bessel and Neumann functions respectively. In (\ref{3.11}) we use the
following notations%
\begin{eqnarray}
\left\{ a\left( x\alpha \right) ,b\left( x\beta \right) \right\} _{x}
&\equiv &a\dfrac{\partial b}{\partial x}-\dfrac{\partial a}{\partial x}b
\notag \\
f_{\underline{l}}\left( x\right) &\equiv &f_{l}\left( x\right) /\left\{
j_{l}\left( xu_{b}\right) ,h_{l}\left( xu\right) \right\} _{x}  \label{3.12}
\\
z_{\nu }^{l} &\equiv &\dfrac{uj_{\nu }\left( xu_{b}\right) h_{l}\left(
xu\right) \varepsilon _{b}-u_{b}j_{l}\left( xu_{b}\right) h_{\nu }\left(
xu\right) }{uj_{\nu }\left( xu_{b}\right) h_{l}\left( xu\right)
-u_{b}j_{l}\left( xu_{b}\right) h_{\nu }\left( xu\right) }.  \notag
\end{eqnarray}%
Remember that the space outside the ball is empty and $x<1$. The derivation
of (\ref{3.9}) is given in \cite{16}.

In case of $x=0$ and/or $\varepsilon _{b}=1$

\begin{equation}
b_{l}(H)=iuj_{l}(u),\qquad b_{l}(E)=i(2l+1)[uj_{l}^{^{\prime }}(u)+j_{l}(u)],
\label{3.13}
\end{equation}%
and therefore, naturally, $n_{k}$ does not depend on $x$. Numerical
calculations by means of formulas (\ref{3.9}) and (\ref{2.10}) give the same
result in case of $\varepsilon _{b}=\varepsilon =1$:%
\begin{equation}
n_{k}\left( \text{ball};\varepsilon _{b}=1\right) =n_{k}\left( \infty
;\varepsilon =1\right) \quad \lbrack \equiv n_{k}\left( \text{vac}\right) ].
\label{3.14}
\end{equation}

\section{ Results of numerical calculations}

\label{section4}

Let us consider the radiation generated by electron at some harmonic $\omega
_{k}=k\omega _{0}$, e.g., at $k=8$.

\begin{figure}[tbph]
\begin{center}
\epsfig{figure=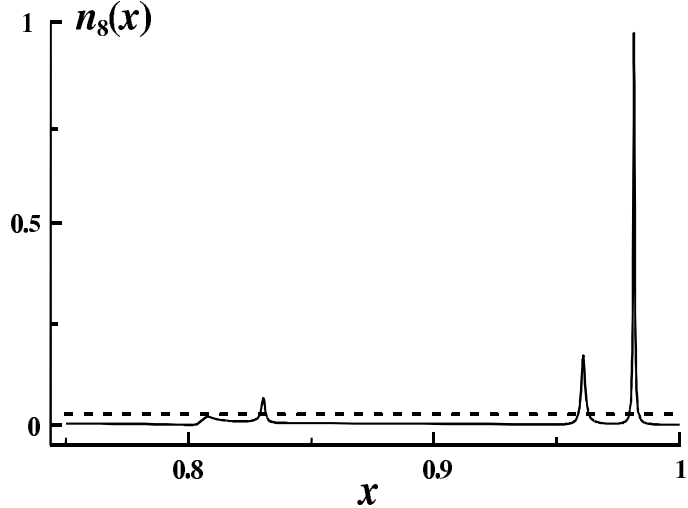,width=9cm,height=6.7cm}
\end{center}
\caption{The number $n_{k}(x)$\ of electromagnetic field quanta generated
per one revolution of electron about a dielectric ball versus the ratio $%
x=r_{b}/r_{e}$ of the ball radius to that of electron orbit. $n_{k}(x)$ is
calculated by formula (\protect\ref{3.9}). The electron energy $E_{e}=2MeV$,
the radius of orbit \ $r_{e}=3.69cm$, the harmonic number $k=8$. The
radiation was emitted respectively at the wavelength of $3cm$ (in vacuum).
The dielectric losses of energy inside the ball material (melted quartz)
were taken into account. For explanations see the text.}
\label{fig1e}
\end{figure}

Plotted in Fig. \ref{fig1e} along the axis of ordinates is the number $n_{8}$%
\ of emitted quanta, and along the abscissa - the radius of ball, or to be
more precise, - the ratio $x=r_{b}/r_{e}$\ of ball-to-electron orbit radii.
The particle energy and radius of particle orbit are fixed at values $%
E_{e}=2MeV$\ and\ $r_{e}=3.69cm$ respectively. For these values of $E_{e}$\
and $r_{e}$, corresponding to the 8-th harmonic is the radiation at
frequency $\omega _{8}/2\pi =10^{10}Hz$ and wavelength $\lambda _{8}=3cm\ $%
(in vacuum). The calculations were carried out by using formula (\ref{3.9})
under the assumption that the ball is made of melted quartz taking into
account the dielectric losses of energy and dispersion of electromagnetic
waves inside the ball material:
\begin{equation}
\varepsilon _{b}=\varepsilon _{b}^{^{\prime }}\left( \omega \right)
+i\varepsilon _{b}^{^{\prime \prime }}\left( \omega \right)  \label{4.1}
\end{equation}%
(for melted quartz $\varepsilon _{b}^{^{\prime }}\left( \omega _{8}\right)
=3.78$, and the dielectric loss angle tangent $\varepsilon _{b}^{^{\prime
\prime }}\left( \omega _{8}\right) /\varepsilon _{b}^{^{\prime }}\left(
\omega _{8}\right) =0.0001$\ \cite{29,30}).

According to (\ref{2.10})), the number of quanta emitted at the 8-th
harmonic by electron with the same values of $E_{e}$\ and $r_{e}$ at the
rotation in a continuous, infinite and transparent medium with $\varepsilon
=3.78$\ would have been $n_{8}\left( \infty \right) \cong 0.0274\sim n_{0}$\
(see the horizontal dashed line in Fig. \ref{fig1e}). It is easy to see \cite%
{7}-\cite{9} that if the electron did not rotate, but moved rectilinearly
with energy $E_{e}=2MeV$\ in the same infinite medium, then during the
period of time $T=2\pi /\omega _{0}$\ it would emit
\begin{equation}
n_{\Delta \omega }\left( \infty \right) =\left( \text{v}/c-c/\text{v}%
\varepsilon \right) n_{0}\cong 0.0318  \label{4.2}
\end{equation}%
quanta in narrow frequency band $\Delta \omega =\omega _{0}$. It is evident
that $n_{\Delta \omega }\left( \infty \right) \sim n_{8}\left( \infty
\right) $. In the absence of ball one has to substitute $x=0$\ (and/or $%
\varepsilon _{b}=1$) in (\ref{3.9}). However, in this case owing to (\ref%
{3.14}) it is simpler to use formula (\ref{2.10}) with $\varepsilon =1$. As
a result $n_{8}\left( \text{vac}\right) \cong 0.00475\approx n_{0}/2k^{2/3}$%
, where $k=8$.

According to data given in Fig. \ref{fig1e}, $n_{8}\left( x\right) \sim
n_{8}\left( \text{vac}\right) \ $practically for all $x$ except for $%
0.8<x<0.85$\ and $0.95<x<1$. In each of these ranges there are peaks. In the
second range the heights of peaks are greater than $n_{8}\left( \text{vac}%
\right) $\ and $n_{8}\left( \infty \right) $\ by many times. For the highest
peak

\begin{equation}
x^{\ast }=0.9815,\text{\quad }n_{8}\left( \text{ball};x^{\ast }\right) \cong
0.951,\quad n_{8}\left( ball;x^{\ast }\right) /n_{8}\left( \infty \right)
\cong 35.  \label{4.3}
\end{equation}%
The corresponding value of ball radius $r_{b}=3.62cm$\ and, consequently,
the distance between the rotating electron and the ball surface should be $%
r_{e}-r_{b}=0.7mm$. Apparently,
\begin{equation}
n_{8}\left( \text{vac}\right) <<n_{\Delta \omega }\left( \infty \right) \sim
n_{8}\left( \infty \right) <<n_{8}\left( \text{ball};x^{\ast }\right) .
\label{4.4}
\end{equation}

Similar results are obtained for a series of other values of $k>>1$, as well
as for electrons with energies $1\leq E_{e}\leq 5MeV$\ and balls with $1\leq
\varepsilon _{b}^{^{\prime }}\leq 5$ and $\varepsilon _{b}^{^{\prime \prime
}}/\varepsilon _{b}^{^{\prime }}<<1$. In all these cases the following
condition
\begin{equation}
\text{v}_{\ast }\sqrt{\varepsilon _{b}^{^{\prime }}}/c>1  \label{4.5}
\end{equation}%
is satisfied. Here v$_{\ast }\equiv x^{\ast }$v$=r_{b}\omega _{0}$. Hence, a
high power radiation with $n_{8}\left( x^{\ast }\right) >>n_{8}\left( \infty
\right) $\ is possible only for particular values of $x=r_{b}/r_{e}$ and
when the Cherenkov condition for velocity of particle "image" on the ball
surface and the real part of the permittivity of ball material is satisfied.

\section{Conclusion}

\label{section5}

In the present paper the radiation from a relativistic electron at uniform
rotation about a dielectric ball has been studied with due regard for
dispersion and dielectric losses of energy inside the ball material. Here,
in addition to the synchrotron radiation the electron may also generate
Cherenkov radiation. Its appearance is attributed to the fact that the
electromagnetic field coupled with electron partially penetrates the ball
and rotates with the particle. In case of small distance between the
relativistic particle and the ball surface, $r_{e}\approx r_{b}$, this field
can propagate with a speed larger than the phase speed of light inside the
ball material, and then Cherenkov radiation is to be generated.

The peculiarities in total radiation at different harmonics due to the
influence of matter and radius of ball have been investigated theoretically.
It was shown that in case of weak absorption ($\varepsilon _{b}^{^{\prime
\prime }}<<\varepsilon _{b}^{^{\prime }}$ ) at some harmonics with $k>>1$,
the electron may generate $n_{k}\approx 1$\ (see (\ref{4.3})) quanta of
electromagnetic field during one rotation period. This value is more than
30-fold greater than the similar value of $n_{k}$\ for an electron rotating
in a continuous, infinite and transparent medium having the same
permittivity as a real part of that for the ball material.

The emission of such a high power radiation is possible only when the ratio
of ball radius to that of an electron orbit takes on a series of fixed
values and the Cherenkov condition (\ref{4.5}) for the speed of particle
"image" on the ball surface and the ball material is satisfied. New
characteristic features and visual explanation of the phenomenon will be
given in forthcoming papers.

\section*{Acknowledgements}

The author is sincerely grateful to Academician A.R. Mkrtchyan for support
and continuous interest to the work. The author has to thank Prof. L.Sh.
Grigoryan very much for setting of problem and guidance during the work. He
is especially thankful to Drs. H.F. Khachatryan and S.R. Arzumanyan for
numerous important discussions and valuable assistance, and to D.Sc. A.A.
Saharian for helpful suggestions and comments.

The work is supported by the Ministry of Education and Science of the
Republic of Armenia (Grant No.0063). 

\end{document}